 \documentclass[twoside,11pt,amsfonts]{article} 
\usepackage{graphicx}
\usepackage{amsmath}
\usepackage{amssymb}
\usepackage{array}

\usepackage{enumerate}
\begin{document}

\begin{center}
{\bf MATRIX-VARIATE STATISTICAL DISTRIBUTIONS
  AND FRACTIONAL CALCULUS}\\

\vskip.3cm\noindent {\bf A.M. Mathai [1], H.J. Haubold [2]}\\

 \vskip.2cm Dedicated to Professor R. Gorenflo on the occasion of his 80th birthday\\
 
\end{center}
\begin{center}
\vskip.5cm {\bf Abstract}
\end{center}
\vskip.3cm A connection between fractional calculus and statistical distribution theory has been established by the authors recently. Some extensions of the results to matrix-variate functions were also considered. In the present article, more results on matrix-variate statistical densities and their connections to fractional calculus will be established. When considering solutions of fractional differential equations, Mittag-Leffler functions and Fox H-function appear naturally. Some results connected with generalized Mittag-Leffler density and their asymptotic behavior will be considered. Reference is made to applications in physics, particularly superstatistics and nonextensive statistical mechanics.



\vskip.5cm\noindent{\bf 1.\hskip.3cm Introduction}

\vskip.3cm
For our discussions to follow we need the definitions of Mittag-Leffler function, Wright function, Fox H-function and generalized Mittag-Leffler density. Hence we will introduce these functions first.
\vskip.2cm
The Mittag-Leffler function is used in many areas of physical sciences [see Haubold, Mathai, and Saxena (2009), Saxena, Mathai, and Haubold (2010), Haubold, Mathai, and Saxena (2011)]. Haubold and Mathai [2000, see also Haubold and Kumar (2010)] have shown that when we move from total differential or integral
equation to fractional cases the exponential type solutions go in terms of Mittag-Leffler functions. For a modern introduction to fractional calculus see Mainardi (2010) and Tenreiro, Machado, Kiryakova, and Mainardi (2011). Solutions of certain fractional reaction-diffusion equations belong to Mittag-Leffler functions [Gorenflo and Mainardi
(1996), Kiryakova (2000), Kilbas (2005), Mainardi and Pagnini (2008), Mathai, Saxena, and Haubold (2010)]. Let us start with the definition of Mittag-Leffler function as proposed by Mittag-Leffler and its various types of
generalizations.

\vskip.3cm\noindent{\bf 1.1.\hskip.3cm Mittag-Leffler Function}

\vskip.3cm The basic Mittag-Leffler function is an extension of the exponential function. The exponential series is given by the following:

\begin{equation*}
\sum_{k=0}^{\infty}\frac{z^k}{k!}=\sum_{k=0}^{\infty}\frac{z^k}{\Gamma(k+1)}
={\rm e}^x.
\end{equation*}
If we make a change in $\Gamma(1+k)$ and write it as
$\Gamma(1+\alpha k)$ for $\Re(\alpha)>0$, where $\Re(\cdot)$ means the real
part of $(\cdot)$ then we have the original Mittag-Leffler function,
namely,
\begin{equation}
\sum_{k=0}^{\infty}\frac{z^k}{\Gamma(1+\alpha k)}=E_{\alpha}(z).
\end{equation}
Thus for $\alpha=1$ we have $E_1(z)={\rm e}^z$. Three forms of Mittag-Leffler functions are frequently
used in the literature and hence we list these three forms first.
\begin{align}
E_{\alpha,\beta}^{\gamma}(x)&=\sum_{k=0}^{\infty}\frac{(\gamma)_k x^k}{k!\Gamma(\beta+\alpha
k)},~(\gamma)_k=\gamma(\gamma+1)...(\gamma+k-1),(\gamma)_0=1,\gamma\ne
0,\alpha>0,\beta>0;\\
E_{\alpha,\beta}^1(x)&=E_{\alpha,\beta}(x)=\sum_{k=0}^{\infty}\frac{x^k}{\Gamma(\beta+\alpha
k)},\alpha>0,\beta>0;\\
E_{\alpha,1}(x)&=E_{\alpha}(x)=\sum_{k=0}^{\infty}\frac{x^k}{\Gamma(1+\alpha
k)},\alpha>0;~~E_1(x)={\rm e}^x.
\end{align}
When the parameters are in the complex domain the conditions become $\Re
(\alpha)>0,~\Re(\beta)>0$. It is easy to note that the Mittag-Leffler
functions of (1)-(4) are special cases of the Wright function [Wright (1940)] which
is defined as
\begin{align}
{_p\psi_q}(z)&={_p\psi_q}(a_1,..,a_p;b_1,...,b_q;z)\nonumber\\
&=\sum_{k=0}^{\infty}\frac{\prod_{j=1}^p\Gamma(a_j+k)}{\prod_{j=1}^q
\Gamma(b_j+k)}\frac{z^k}{k!}
\end{align}
where the $a_j$'s and $b_j$'s are complex parameters. This Wright function is
a special case of the most generalized special function, namely, the
H-function and the H-function is defined by the following Mellin-Barnes
integral:
\begin{equation}
H_{p,q}^{m,n}(z) = H^{m,n}_{p,q}\left[z\big\vert_{(b_1,\beta_1),...,
(b_q,\beta_q)}^{(a_1,\alpha_1),...,(a_p,\alpha_p)}\right]
=\frac{1}{2\pi i}\int_{c-i\infty}^{c+i\infty}\phi(s)z^{-s}{\rm d}s
\end{equation}
where
\begin{equation}
\phi(s) =\frac{\{\prod_{j=1}^m\Gamma(b_j+\beta_js)\}
\{\prod_{j=1}^n\Gamma(1-a_j-\alpha_js)\}}{\{\prod_{j=m+1}^q\Gamma(1-b_j
-\beta_js)\}
\{\prod_{j=n+1}^p\Gamma(a_j+\alpha_js)\}}
\end{equation}
where $a_j$'s and $b_j$'s are complex quantities, $\alpha_j$'s and
$\beta_j$'s are positive real numbers, $i=\sqrt{-1}$. Various types of
possible contours, convergence conditions and applications may be seen from
Mathai, Saxena and Haubold (2010). The elementary special
functions such as binomial function, Bessel functions of various types, sine
and cosine functions, hypergeometric functions of different types etc, which
are applicable in physical sciences, are
special cases of (6).

\vskip.5cm\noindent{\bf 1.2.\hskip.3cm A Generalized Mittag-Leffler Statistical Density}

\vskip.3cm A statistical density in terms of Mittag-Leffler function was originally defined by Pillai (1990) in terms of the following distribution function or cumulative density:
\begin{equation}
G_y(y)=1-E_{\alpha}(-y^{\alpha})=\sum_{k=1}^{\infty}\frac{(-1)^{k+1}y^{\alpha
k}}{\Gamma(1+\alpha k)},~0<\alpha\le 1,~ y>0
\end{equation}
and $G_y(y)=0$ for $y\le 0$. Since it is differentiable one can obtain the density function by differentiating on both sides with
respect to $y$ and the density function $f(y)$ is the following:
\begin{align}
f(y)&=\frac{{\rm d}}{{\rm d}y}G_y(y)\nonumber\\
&=\frac{{\rm d}}{{\rm
d}y}\left[\sum_{k=1}^{\infty}\frac{(-1)^{k+1}y^{\alpha
k}}{\Gamma(1+\alpha
k)}\right]=\sum_{k=1}^{\infty}\frac{(-1)^{k+1}\alpha k y^{\alpha
k-1}}{\Gamma(1+\alpha k)}\nonumber\\
&=\sum_{k=1}^{\infty}\frac{(-1)^{k+1}y^{\alpha k-1}}{\Gamma(\alpha
k)}=\sum_{k=0}^{\infty}\frac{(-1)^ky^{\alpha+\alpha
k-1}}{\Gamma(\alpha+\alpha k)}
\end{align}
by replacing
$k$ by $k+1$
\begin{equation}
f(y)=y^{\alpha-1}E_{\alpha,\alpha}(-y^{\alpha}),~0<\alpha\le 1,
y>0
\end{equation}
where $E_{\alpha,\beta}(x)$ is the generalized Mittag-Leffler
function.
\vskip.2cm In order to study the properties of the Mittag-Leffler function, the
corresponding fractional differential or integral equations and their
generalizations, usually the techniques from Laplace and Fourier transforms
are used. Hence we will consider some Laplace transforms and then we will establish some connections to
L\'evy distributions. The Laplace transform of (10) is
\begin{align}
L_f(t)&=\int_0^{\infty}{\rm e}^{-tx}f(x){\rm
d}x=\int_0^{\infty}{\rm
e}^{-tx}x^{\alpha-1}E_{\alpha,\alpha}(-x^{\alpha}){\rm d}x\\
&=(1+t^{\alpha})^{-1}, |t^{\alpha}|<1.
\end{align}
Note that (12) is a special case of the general class of Laplace
transforms discussed in Section 2.3.7 of Mathai and Haubold (2008). From (12)
one can also note that there is a structural representation in terms of positive L\'evy distribution.

\vskip.3cm\noindent{\bf Definition 1.1.}\hskip.3cm A positive L\'evy random variable $u>0$, with
parameter $\alpha$ is such that the Laplace transform of the density
of $u>0$ is given by ${\rm e}^{-t^{\alpha}}$. That is,
\begin{equation}
E[{\rm e}^{-tu}]={\rm e}^{-t^{\alpha}}
\end{equation}
where $E(\cdot)$ denotes statistical expectation or the  expected value of
$(\cdot)$.

\vskip.3cm When
$\alpha=1$ the random variable is degenerate with the density/
probability function
\begin{equation}
f_1(x)=\begin{cases}1, \hbox{ for  } x=1\\
0,\mbox{  elsewhere.}
\end{cases}
\end{equation}
Consider a gamma random variable with the scale parameter $\delta$
 and shape parameter $\beta$ or with the density function
\begin{equation}
f_1(x)=\begin{cases}\frac{x^{\beta-1}{\rm
e}^{-\frac{x}{\delta}}}{\delta^{\beta}\Gamma(\beta)},
\mbox{  for  }0\le x<\infty,\beta>0,\delta>0\\
 0,\mbox{
elsewhere;}
\end{cases}
\end{equation}
and with the Laplace transform
\begin{equation}
L_{f_1}(t)=(1+\delta t)^{-\beta}.
\end{equation}
\vskip.2cm A more general form of the Mittag-Leffler  density is given by the following [Mathai (2010)]:

\begin{align}
g(x)&=\sum_{k=0}^{\infty}\frac{(-1)^k(\eta)_k}{k!\Gamma(\alpha\eta+\alpha k)}\frac{x^{\alpha\eta-1+\alpha k}}{a^{\eta+k}}\\
&=\frac{x^{\alpha\eta-1}}{a^{\eta}}E_{\alpha,\alpha\eta}^{\eta}\left(-\frac{x^{\alpha}}{a}\right)
\end{align}
and with the Laplace transform
\begin{equation}
L_g(t)=[1+at^{\alpha}]^{-\eta}, ~|at^{\alpha}|<1
\end{equation}
where $a>0,~0<\alpha\le 1,~\eta>0$.

With these preliminaries we will discuss a few problems connected with Mittag-Leffler density and we will examine the matrix-variate analogues.

\vskip.3cm\noindent{\bf 2.\hskip.3cm Generalized Mittag-Leffler Density}

\vskip.3cm First, we will examine a minor generalization of a known result, which is given recently by Mathai
(2010). The proof is given by others by using different methods but we will obtain it by using
some properties of conditional expectations. To this end, we will list a basic result on conditional expectations in the form of the following lemma.

\vskip.3cm \noindent{\bf Lemma 2.1.}\hskip.3cm{\it For two random
variables $u$ and $v$ having a joint distribution,
\begin{equation}
E(u)=E[E(u|v)]
\end{equation}
whenever all the expected values exist, where the inside
expectation is taken in the conditional space of $u$ given $v$ and
the outside expectation is taken in the marginal space of $v$.}

\vskip.3cm A real scalar random variable $x$ is a mathematical variable where a probability statement of the type $Pr\{x\le \alpha\}$, that is the probability of $x$ falling in the interval $-\infty<x\le \alpha$, is defined for all real values of $\alpha$. A density function $f(x)$, associated with the random variable $x$, is a non-negative integrable function with the total integral unity. The expected value of a function $\phi(x)$ of $x$, denoted by $E[\phi(x)]$, is the integral
\begin{equation}
E[\phi(x)]=\int_{-\infty}^{\infty}\phi(x)f(x){\rm d}x
\end{equation}
where $f(x)$ is the density function of $x$ when $x$ has nonzero probabilities on a continuum of points. Parallel definitions go for discrete random variables, joint densities of many random variables, vector or matrix random variables etc.

\vskip.2cm The lemma follows from the definition itself. By using this lemma and equations (13), (15) and (17) one can obtain a structural representation for a Mittag-Leffler random variable. This will be stated as a theorem.

\vskip.3cm \noindent{\bf Theorem 2.1.}\hskip.3cm{\it Let $y>0$ be a
L\'evy random variable with Laplace transform as in (13), $x\ge 0$
be a gamma random variable with the density as in (15) and let $x$
and $y$ be independently distributed. Then $u=yx^{\frac{1}{\alpha}}$
is distributed as a Mittag-Leffler random variable with Laplace
transform
\begin{equation}
L_u(t)=[1+\delta t^{\alpha}]^{-\beta}.
\end{equation}
}
\vskip.3cm \noindent{\bf Proof}:\hskip.3cm For proving this result
we will make use of the conditional argument as given in Lemma 2.1.
Let the
density of $u$ be denoted by $g(u)$. Then the conditional Laplace transform of
$g$, given $x$, is given by
\begin{equation}
E[{\rm e}^{-(tx^{\frac{1}{\alpha}})y}|x]={\rm
e}^{-t^{\alpha}x}.
\end{equation}
But the right side of (23) is in the form of a Laplace
transform of the density of $x$ with parameter $t^{\alpha}$. Hence
the expected value of the right side, with respect to $x$, is available by replacing the Laplace parameter $t$ in (16) by $t^{\alpha}$,
\begin{equation}
L_g(t)=(1+\delta t^{\alpha})^{-\beta},\delta>0,\beta>0
\end{equation}
which establishes the result. From (20) one property is
obvious. Suppose that we consider an arbitrary random variable $y$
with the Laplace transform of the form
\begin{equation}
L_g(t)={\rm e}^{-[\phi(t)]}
\end{equation}
whenever the expected value exists, where $\phi(t)$ be such that

\begin{equation}
\phi(tx^{\frac{1}{\alpha}})=x\phi(t), \lim_{t\rightarrow
0}\phi(t)=0.
\end{equation}

Then from (23) we have
\begin{equation}
E[{\rm e}^{-(tx^{\frac{1}{\alpha}})y}|x]={\rm
e}^{-x[\phi(t)]}.
\end{equation}
Now, let $x$ be an arbitrary positive
random variable having Laplace transform, denoted by $L_x(t)$ where
$L_x(t)=\psi(t)$. Then from (25) and (20) we have
\begin{equation}
L_g(t)=\psi[\phi(t)].
\end{equation}

\vskip.2cm General properties of Mittag-Leffler functions, their generalizations, their applications to fractional calculus, reaction-diffusion type problems, may be seen from papers, for example, Gorenflo and Mainardi (1996), Gorenflo, Kilbas, and Rogosin (1998), Gorenflo, Loutchko, and Luchko (2002), Kilbas (2005), and Kilbas and Saigo (1996).

\vskip.2cm We can create a generalized Mittag-Leffler density as follows: Consider the generalized Mittag-Leffler function

\begin{align}
g_1(x)&=\frac{1}{\Gamma(\eta)}\sum_{k=0}^{\infty}\frac{(-1)^k
\Gamma(\eta+k)}{k!\Gamma(\alpha
k+\alpha \eta)}x^{\alpha\eta-1+\alpha k}\\
&=x^{\alpha\eta-1}E_{\alpha\eta,\alpha}^{\eta}(-x^{\alpha}),\alpha>0,\eta>0.
\end{align}
Then the Laplace transform of $g_1(x)$ is the following:

\begin{align}
L_{g_1}(t)&=\sum_{k=0}^{\infty}\frac{(-1)^k}{k!}\frac{(\eta)_k}
{\Gamma(\alpha\eta+\alpha
k)}\int_0^{\infty}x^{\alpha\eta+\alpha k-1}{\rm e}^{-tx}{\rm
d}x\\
 &=\sum_{k=0}^{\infty}(-1)^k\frac{(\eta)_k}{k!}t^{-\alpha
\eta-\alpha k}=(1+t^{\alpha})^{-\eta},|t^{\alpha}|<1.
\end{align}
This is a special case of a general class of Laplace transforms considered in
Mathai, Saxena, and Haubold (2006), Saxena, Mathai, and Haubold (2010), and Haubold, Mathai, and Saxena (2011).

\vskip.3cm\noindent{\bf 2.1.\hskip.3cm Mellin-Barnes Representations}

 \vskip.3cm It is shown in Mathai (2010) that for handling problems connected with Mittag-Leffler densities it is convenient to use the Mellin-Barnes representation of a Mittag-Leffler function and then proceed from there.  $g_1(x)$ of (30) can be written as a Mellin-Barnes
integral and then as an H-function.
Then
\begin{align}
g_1(x)&=\frac{1}{\Gamma(\eta)}\frac{1}{2\pi
i}\int_{c-i\infty}^{c+i\infty}\frac{\Gamma(s)\Gamma(\eta-s)}{\Gamma(\alpha\eta-\alpha
s)}x^{\alpha\eta-1}(x^{\alpha})^{-s}{\rm
d}s,~0<c<\Re(\eta)\\
&=\frac{x^{\alpha\eta-1}}{\Gamma(\eta)}H_{1,2}^{1,1}\left[x^{\alpha}
\big\vert_{(0,1),(1-\alpha\eta,\alpha)}^{(1-\eta,1)}\right]\\
&=\frac{1}{\alpha\Gamma(\eta)}\frac{1}{2\pi
i}\int_{c_1-i\infty}^{c_1+i\infty}
\frac{\Gamma(\eta-\frac{1}{\alpha}+\frac{s}{\alpha})\Gamma(\frac{1}
{\alpha}-\frac{s}{\alpha})}
{\Gamma(1-s)}x^{-s}{\rm d}s
\end{align}
by taking
$\alpha\eta-1-\alpha s=-s_1$,

\begin{align}
g_1(x)&=\frac{1}{\alpha\Gamma(\eta)}H_{1,2}^{1,1}\left[x\big\vert_{(\eta-\frac{1}
{\alpha},
\frac{1}{\alpha}),(0,1)}^{(1-\frac{1}{\alpha},\frac{1}{\alpha})}\right].
\end{align}
One can also look upon the Mellin transform of the density of a positive random variable as  the $(s-1)$-th
moment and therefore, from (2.15) one can write the Mellin transform as an expected value. That is,

\begin{align}
M_{g_1}(s)&=E(x^{s-1})\mbox{  in  }g_1\\
&=\frac{1}{\Gamma(\eta)}\frac{\Gamma(\eta-
\frac{1}{\alpha}+\frac{s}{\alpha})\Gamma(\frac{1}{\alpha}-\frac{s}{\alpha})}
{\alpha\Gamma(1-s)},\\
&=\frac{1}{\Gamma(\eta)}\frac{\Gamma(\eta-
\frac{1}{\alpha}+\frac{s}{\alpha})
\Gamma(1+\frac{1}{\alpha}-\frac{s}{\alpha})}
{\Gamma(2-s)},
\end{align}
for
$1-\alpha<\Re(s)<1,~0<\alpha\le 1,~ \eta>0.$ Observe that the right side in (39) goes to $1$ when $s\rightarrow 1$ thereby establishing that the non-negative function $g_1(x)$ is a density function. This is the most convenient way of showing that $g_1(x)$ is a density because term by term integration from $0$ to $\infty$ is not possible in a Mittag-Leffler function, which is in the form of a  series. This $g_1(x)$  is called the generalized Mittag-Leffler density.

\vskip.2cm From (36) we may also observe the following properties. If $u$ is the generalized Mittag-Leffler variable then from (35) and (36) we have

\begin{align}
E(u^{s-1})&=\frac{1}{\Gamma(\eta)}\frac{\Gamma(\eta-
\frac{1}{\alpha}+\frac{s}{\alpha})
\Gamma(1+\frac{1}{\alpha}-\frac{s}{\alpha})}
{\Gamma(2-s)}\\
&=E(y^{s-1})E(x^{\frac{1}{\alpha}})^{s-1}
\end{align}
where

\begin{equation}
E(y^{s-1})=\frac{\Gamma(1+\frac{1}{\alpha}-\frac{s}{\alpha})}{\Gamma(2-s)}
\end{equation}
and
\begin{equation}
E(x^{\frac{1}{\alpha}})^{s-1}=\frac{1}{\Gamma(\eta)}\Gamma(\eta-\frac{1}{\alpha}+\frac{s}{\alpha})
\end{equation}
which suggests that the Mittag-Leffler random variable has the representation
\begin{equation}
u=y(x^{\frac{1}{\alpha}})
\end{equation}
where $y$ is a positive L\'evy random variable with the $(s-1)$-th moment in (42), $x$ is a gamma random variable with the density function in (15), $u$ is a Mittag-Leffler variable and it is assumed that $x$ and $y$ are statistically independently distributed. From (42) it may be further noted that $y$ does not have a density but for large values of $y$ one can have an asymptotic form of the density of $y$.

\vskip.5cm\noindent{\bf 3.\hskip.3cm The Pathway Model}

\vskip.3cm In model building situations in physical sciences, what is usually done is to select a model from a parametric family of distributions if it is a deterministic (non-random) or random  (non-deterministic) situation, such as a gamma family of distributions. A gamma family of distributions usually has two parameters $(\alpha,\beta)$. Thus these two parameters are suitably chosen and then a model is selected for the data at hand. Often it is found that the selected model is not a good fit because the data requires a model with a thicker or thinner tail than the one given by the parametric family, or it may be that the appropriate model is in between two parametric families of distributions. In order to come up with an appropriate model in such situations a pathway model was proposed in Mathai [2005, see also Mathai and Haubold (2010), Mathai, Haubold, and Tsallis (2010)].
The original pathway model of Mathai (2005) is for the rectangular matrix cases. The scalar version of the pathway model is the following:

\begin{equation}
f_1(x)=c_1|x|^{\gamma}[1-a(1-\alpha)|x|^{\delta}]^{\frac{\eta}{1-\alpha}}
\end{equation}
for $a>0,~\eta>0,~\delta>0,~\gamma>-1$ and $c_1$ is a constant. A statistical density can be created out of $f_1(x)$ under the additional conditions $1-a(1-\alpha)|x|^{\delta}>0$ and then $c_1$ will act as a normalizing constant. Hence the model is directly applicable to deterministic situations as well as random or non-deterministic situations. For $\alpha<1$ the model in (45) will stay in the generalized type-1 beta family of functions. For $\alpha>1$, writing $1-\alpha=-(\alpha-1)$ the model switches into the model

\begin{equation}
f_2(x)=c_2|x|^{\gamma}[1+a(\alpha-1)|x|^{\delta}]^{-\frac{\eta}{\alpha-1}}
\end{equation}
for $\alpha>1$, $-\infty<x<\infty$, $a>0,~\eta>0,\delta>0$. Note that when $\alpha $ switches from $\alpha<1$ to $\alpha >1$ the model goes from a generalized type-1 beta family of functions to a generalized type-2 beta family of functions. Again, taking $c_2$ as the normalizing constant one can create a statistical density out of $f_2(x)$. When $\alpha$ goes to $1$ from the right or from the left, $f_1(x)$ and $f_2(x)$ will go into a generalized gamma family of functions, namely,

\begin{equation}
\lim_{\alpha\rightarrow 1_{-}}f_1(x)=\lim_{\alpha\rightarrow 1_{+}}f_2(x)=f_3(x)=c_3|x|^{\gamma}{\rm e}^{-a\eta|x|^{\delta}}
\end{equation}
where $a>0,~\eta>0,~\delta>0$. It is shown in Mathai and Haubold (2007) that almost all statistical densities in current use in statistics and physics are special cases of the pathway model of (45) to (47) or as a one-to-one function of the variable therein. The matrix-variate pathway model is of the following form:

\begin{align}
f(X)&=c|A^{\frac{1}{2}}(X-M)B(X-M)'A^{\frac{1}{2}}|^{\gamma}\nonumber\\
&\times  |I-a(1-\alpha)A^{\frac{1}{2}}(X-M)B(X-M)'A^{\frac{1}{2}}|^{\frac{\eta}{1-\alpha}}
\end{align}
where a prime denotes a transpose,  $M,A,B$ are constant matrices, $M$ is $m\times n$,$~m\le n$, $A=A'>0$ is $m\times m$ and real positive definite, $B=B'>0$ is $n\times n$ and real positive definite, $X$ is $m\times n$ matrix of distinct real variables and of rank $m$, and $|(\cdot)|$ denotes the determinant of $(\cdot)$. Note that for $m=1$ one has a quadratic form in pathway vectors and then (48) can produce an extended density of quadratic forms, see Mathai (2007).

\vskip.2cm One can establish a path of going from a Mittag-Leffler variable to a positive L\'evy variable. Replace $a$ by $a(q-1)$ and $\eta$ by $\eta/(q-1)$ in the Laplace transform (22) of the generalized Mittag-Leffler density in (8) and (18). That is,

\begin{equation}
L_g(t)=[1+a(q-1)t^{\alpha}]^{-\frac{\eta}{q-1}},
q>1.
\end{equation}
If $q\rightarrow 1_{+}$ then
\begin{equation}
L_g(t)\rightarrow {\rm e}^{-a\eta t^{\alpha}}=L_{u}(t)
\end{equation}
which is the Laplace transform of a constant multiple of a positive
L\'evy variable with parameter $\alpha$, given in (13). Thus $q$ here creates a
pathway of going from the general Mittag-Leffler density $g$ to a
positive L\'evy density  with parameter $\alpha$, the
multiplying constant being $(a\eta)^{\frac{1}{\alpha}}$. For a
discussion of a general rectangular matrix-variate pathway model see
Mathai (2005). The result in (48) can be put in a more general
setting also. These ideas are generalized to the matrix-variate cases recently, see Mathai (2010). Some multivariable, not matrix variable, analogues may be seen from Lim and Teo (2009) and Pakes (1998).

\vskip.2cm Another area of applications of Mittag-Leffler functions, Mittag-Leffler density and pathway models is in the area of quadratic and bilinear forms in Gaussian variables. The theory and applications may be seen from the books of Mathai and Provost (1992) and Mathai, Provost and Hayakawa (1995). With the help of the pathway models of (45) to (47) one can extend the theory of quadratic and bilinear forms to wider classes, rather than confining to Gaussian variables. Some works in matrix variate cases in this direction may be seen from Mathai (2007).

\vskip.2cm Particular cases of the models in (45) to (47) are Tsallis statistics and superstatistics of non-extensive statistical mechanics. In (45) and (46) put $\gamma=0,~\delta=1,~a=1,~\eta=1$ and consider only positive variable $x>0$ then we get the statistics of Tsallis (1988, 2004, 2009). Superstatistics is not available from the form in (45) for $\alpha<1$. Consider (46) for $x>0$. Put $a=1,~\delta=1,~\eta=1$ then we get the superstatistics of Beck and Cohen (2003) and Beck (2006, 2008, 2009, 2010).

\vskip.2cm The models in (45) to (47) can also be derived by optimizing a generalized measure of entropy under some moment-type conditions, see Mathai and Haubold (2007).

\vskip.3cm \noindent{\bf 4.\hskip.3cm Multivariable Analogues}

\vskip.3cm A slight extension of the L\'evy variable is the Linnik variable. A Linnik random variable is defined as that real scalar
random variable whose characteristic function is given by

\begin{equation}
\phi(t)=\frac{1}{1+|t|^{\alpha}},~0<\alpha\le
2,~-\infty<t<\infty.
\end{equation}
For $\alpha=2$, (51) corresponds to the characteristic function
of a Laplace random variable and hence Pillai and his coworkers name
the distribution corresponding to (51) as the $\alpha$-Laplace
distribution.  Infinite
divisibility, characterizations, other properties and related
materials may be seen from the review paper Jayakumar and Suresh
(2003) and the references therein, Pakes (1998) and Mainardi
and Pagnini (2008). Multivariate generalization of Mittag-Leffler
and Linnik distributions may be seen from Lim and Teo (2009). Since
the steps for deriving results on Linnik distribution are parallel
to those of the Mittag-Leffler variable, further discussion of
Linnik distribution is omitted. The $\alpha$-Laplace
distribution plays vital roles in non-Gaussian stochastic processes
and time series.

\vskip.2cm A multivariate Linnik distribution can be defined in
terms of a multivariate L\'evy vector. Let $T'=(t_1,...,t_p),
X'=(x_1,...,x_p)$, prime denoting the transpose. A vector variable
having positive L\'evy distribution is given by the characteristic
function

\begin{equation}
E[{\rm e}^{iT'X}]={\rm e}^{-(T'\Sigma T)^{\frac{\alpha}{2}}},~
0<\alpha\le 2,
\end{equation}
where $\Sigma=\Sigma'>0$ is a real positive definite $p\times p$
matrix. We consider a representation, corresponding to the one in (44).

\begin{equation}
U=y^{\frac{1}{\alpha}}X
\end{equation}
where the $p\times 1$ vector $X$, having a multivariable L\'evy
distribution with parameter $\alpha$, and $y$ a real scalar gamma
random variable with the parameters $\delta$ and $\beta$, are
independently distributed. Then the characteristic function of the
random vector variable $U$ is given by the following:

\begin{align}
E[{\rm e}^{iy^{\frac{1}{\alpha}}T'X}]&=E[E[{\rm
e}^{iy^{\frac{1}{\alpha}}T'X}]|_{y}]\\
 &=E[{\rm e}^{-y[(T'\Sigma
T)]^{\frac{\alpha}{2}}}]=[1+\delta (T'\Sigma
T)^{\frac{\alpha}{2}}]^{-\beta}.
\end{align}
Then the distribution of $U$, with the characteristic function in
(55) is called a {\it vector-variable Linnik distribution}. Some
properties of this distribution are given in Lim and Teo (2009).

\vskip.3cm  One can
also establish some central limit property  but it will be
of the nature of generalized Mittag-Leffler variable going to a
positive L\'evy variable. Consider the generalized Mittag-Leffler
density with the Laplace transform

\begin{equation}
L_x(t)=[1+\delta t^{\alpha}]^{-\beta},~ \delta>0,~\beta>0,~
0<\alpha\le 1.
\end{equation}
Let $x_j,j=1,...,n$ be independently and identically distributed
as in (56) and let

\begin{equation}
y=\frac{(x_1+...+x_n)}{n^{\frac{1}{\alpha}}}
\end{equation}
Then

\begin{equation}
L_y(t)=[1+\frac{\delta t^{\alpha}}{n}]^{-n\beta}.
\end{equation}
Hence when $n\rightarrow\infty$

\begin{equation}
\lim_{n\rightarrow\infty}L_y(t)={\rm e}^{-\delta\beta
t^{\alpha}}
\end{equation}
which is the Laplace transform of a constant multiple of a
positive L\'evy variable with parameter $\alpha$, the constant being
$(\delta\beta)^{\frac{1}{\alpha}}$. Thus the central limiting
property is that a certain normalized sample mean from a
Mittag-Leffler population goes to a L\'evy variable. One can also introduce the pathway model here. If we replace $\delta$ by $\delta (q-1)$ and $\beta$ by
$\frac{\beta}{q-1},~ q>1$ then as $q\rightarrow 1$,

\begin{equation}
\lim_{q\rightarrow
1}[1+\delta(q-1)t^{\alpha}]^{-\frac{\beta}{q-1}}={\rm
e}^{-\delta\beta t^{\alpha}}
\end{equation}
where $q$ is a pathway parameter describing the path of going from
a general Mittag-Leffler variable to a constant multiple of a L\'evy
variable.

\vskip.2cm We can extend the above ideas to matrix-variate cases also. Let $X=(x_{rs})$ be $m\times n$ where all $x_{rs}$'s are
distinct, $X$ be of full rank, and having a joint density $f(X)$, where $f(X)$ is a real-valued scalar function of $X$.
Then the characteristic function of $f(X)$, denoted by $\phi_X(T)$,
is given by

\begin{equation}
\phi_X(T)=E[{\rm e}^{i{\rm tr}(XT)}],i=\sqrt{-1}\mbox{  and
}T=(t_{rs})
\end{equation}
is an $n\times m$ matrix of distinct parameters $t_{rs}$'s and let
$T$ be of full rank. As an example, we can look at the real matrix-variate Gaussian distribution, given by the density

\begin{equation}
g(X)=\frac{|A|^{\frac{n}{2}}|B|^{\frac{m}{2}}}{\pi^{\frac{mn}{2}}}{\rm
e}^{-{\rm tr}(AXBX')}
\end{equation}
where $X$ is $m\times n$, $A=A'>0, B=B'>0$ are $m\times m$ and
$n\times n$ positive definite constant matrices, a prime denoting a
transpose. Then the Fourier transform of $g(X)$ is given by the
following:

\begin{align}
\phi_g(T)&=E[{\rm e}^{i{\rm tr}(XT)}], i=\sqrt{-1}\\
&={\rm e}^{-\frac{1}{4}{\rm
tr}(B^{-\frac{1}{2}}TA^{-1}T'B^{-\frac{1}{2}})}
\end{align}
where $T$ is the parameter matrix, and ${\rm tr}(\cdot)$ denotes the trace of $(\cdot)$. Motivated by (61) and (64), Mathai (2010) defined matrix-variate Linnik density and gamma-Linnik density. We will list some of them here.

\vskip.3cm \noindent{\bf Definition 4.1.\hskip.3cm Matrix-variate
Linnik density}.\hskip.3cm Let $X$ and $T$ be as defined in
(61). Then $X$ will be called a real rectangular matrix-variable
Linnik random variable if its characteristic function is given by

\begin{equation}
\phi_X(T)={\rm e}^{-[{\rm
tr}(T\Sigma_1T'\Sigma_2)]^{\frac{\alpha}{2}}},~0<\alpha\le 2
\end{equation}
where $\Sigma_1>0$ is $m\times m$ and $\Sigma_2>0$ is $n\times n$
positive definite constant matrices.

\vskip.2cm For the real rectangular matrix-variate Gaussian density
of (62), $\Sigma_1=\frac{1}{2}A^{-1}$ and
$\Sigma_2=\frac{1}{2}B^{-1}$ and $\alpha=2$. The following result is given recently in Mathai (2010).

\vskip.3cm \noindent{\bf Theorem 4.1.}\hskip.3cm{\it Let $y$ be a real
scalar random variable, distributed independently of the $m\times n$
matrix $X$ and having a gamma density with the Laplace transform

\begin{equation}
L_y(t)=(1+\delta t)^{-\beta},~ \delta>0,~\beta>0
\end{equation}
and let $X$ be distributed as a real rectangular matrix-variate
Linnik variable as given in Definition 4.1 then the rectangular matrix
$U=y^{\frac{1}{\alpha}}X$ has the characteristic function

\begin{equation}
\phi_U(T)=[1+\delta({\rm
tr}(T\Sigma_1T'\Sigma_2))^{\frac{\alpha}{2}}]^{-\beta}.
\end{equation}
}\vskip.3cm As special cases we have the following corollaries:

\vskip.3cm \noindent{\bf Corollary 4.1}\hskip.3cm{\it When
$m=1,\Sigma_1=I_1$ the characteristic function in (67) reduces to

\begin{equation}
\phi_U(T)=[1+\delta(T'\Sigma_2T)^{\frac{\alpha}{2}}]^{-\beta}
\end{equation}
which is the characteristic function of the multivariable Linnik
variable with parameters $\delta,\beta,\Sigma_2>0$ defined in Lim
and Teo (2009).}

\vskip.3cm \noindent{\bf Corollary 4.2.}\hskip.3cm{\it When
$n=1,\Sigma_2=I_1$ then again $\phi_U(T)$ is the same as in
Corollary 4.1 with the parameters $\delta,\beta,\Sigma_1>0$.}

\vskip.3cm \noindent{\bf Corollary 4.3.}\hskip.3cm{\it When
$m=1,\Sigma_1=1,\Sigma_2={\rm diag}(\sigma_{11},..,\sigma_{nn}),
\sigma_{jj}>0, j=1,...,n$ then $\phi_U(T)$ of (67) is given by

\begin{equation}
\phi_U(T)=[1+\delta(\sigma_{11}t_1+...
+\sigma_{nn}t_n)^{\frac{\alpha}{2}}]^{-\beta}.
\end{equation}
}\vskip.3cm Recently Mathai (2010) defined a few matrix random variables in the categories of Gaussian Linnik, Gamma Linnik and Kiryakova Linnik. Some of these will be given here

\vskip.3cm \noindent{\bf Definition 4.2.}\hskip.3cm The matrix
variable $U$ in (67) will be called a real rectangular
matrix-variate Gaussian Linnik variable when $\alpha=2$ and Gamma
Linnik variable for the general $\alpha$.

\vskip.2cm By using the multi-index Mittag-Leffler function of
Kiryakova (2000) one can also define a Gamma-Kiryakova vector or
matrix variable.

\vskip.2cm Now, let us consider two independently distributed real
random variables $y$ and $X$ where $y$ has a general Mittag-Leffler
density and $X$ has a real rectangular matrix-variate Gaussian
density as in (62). Let the density of $y$ be given by

\begin{equation}
f_y(y)=\frac{y^{\alpha\beta-1}}{(\delta)^{\beta}}
\sum_{k=0}^{\infty}\frac{(-1)^k(\beta)_k}{k!(\delta)^k}\frac{y^{\alpha
k}}{\Gamma(\alpha\beta+\alpha k)},~ y\ge
0,~\delta>0,~\beta>0.
\end{equation}
Then the following are the results given recently in Mathai (2010).

\vskip.3cm \noindent{Theorem 4.2.}\hskip.3cm{\it Let $y$ have the
density in (70) and the real rectangular matrix $X$ have the
density in (62) and let $X$ and $y$ be independently distributed.
Then $U_1=y^{\frac{1}{\alpha}}X$ has the characteristic function

\begin{equation}
\phi_{U_1}(T)=[1+\frac{\delta}{4^{\alpha}}({\rm
tr}(A^{-1}T'B^{-1}T))^{\alpha}]^{-\beta}.
\end{equation}
}

\vskip.3cm \noindent{Theorem 4.3.}\hskip.3cm{\it Let
$U_2=y^{\frac{1}{\gamma}}X$ where $y$ and $X$ are independently
distributed with $y$ having the density in (70) and $X$ is real
rectangular matrix-variate Linnik variable with the parameters
$\gamma,A,B$ then $U_2$ has the characteristic function

\begin{equation}
\phi_{U_2}(T)=[1+\delta({\rm
tr}(A^{-1}T'B^{-1}T))^{\frac{\alpha\gamma}{2}}]^{-\beta},~\delta>0,~
\gamma>0,~\beta>0,~\alpha>0.
\end{equation}
}

\vskip.2cm \noindent This $U_2$ will be called a L\'evy
Mittag-Leffler real rectangular matrix-variate random variable.

\vskip.3cm \noindent {\bf 5. Concluding Remarks} \vskip.3cm The Mittag-Leffler functions and their generalizations to vector and matrix variable cases, discussed in this paper will be useful for investigators in disciplines of physical sciences, particulalrly superstatistics [Beck (2006, 2008, 2009, 2010)] and nonextensive statistical mechanics [Tsallis (1988, 2004, 2009)]. The importance of Mittag-Leffler function in physics is steadily increasing. It is simply said that deviations of physical phenomena from exponential behavior could be governed by physical laws through Mittag-Leffler functions (power-law). The paper also continues to show that special functions like Fox H-function and Mittag-Leffler function, represented as Mellin-Barnes integrals of the product of gamma functions are suited to represent statistics of products and quotients of independent random variables [Cottone, Di Paola, and Metzler (2010)]. In this regard, Mathai's (2005) pathway model [see also Mathai and Haubold (2010), Mathai, Haubold, and Tsallis (2010)] allows to switch from a generalized type-1 beta family of functions to a generalized type-2 beta family of functions and further to a generalized gamma family of functions.

\vskip.5cm \noindent{\bf Acknowledgment}

\vskip.3cm The authors would like to thank the Department of Science and Technology, Government of India for the financial assistance for this work under project number SR/S4/MS: 287/05.

\vskip.5cm \noindent
\begin{center}{\bf References}
\end{center}

\vskip.2cm \noindent [1] C. Beck, Stretched exponentials from superstatistics, Physica A, 365(2006), 56-101;

\vskip.2cm \noindent [2] C. Beck, in R.Klages, G. Radons, and I.M. Sokolov: Anomalous Transport: Foundations and Applications, Wiley-VCH, Weinheim, 2008, pp. 433-457;

\vskip.2cm \noindent [3] C. Beck, Recent developments in superstatistics, Brazilian Journal of Physics, 39(2009), 357-363;

\vskip.2cm \noindent [4] C. Beck, Generalized statistical mechanics for superstatistical systems, arXiv.org/1007.0903 [cond-mat.stat-mech], 6 July 2010.

\vskip.2cm \noindent [5] C. Beck and E.G.D. Cohen, Superstatistics, Physica A, 322(2003), 267-275.

\vskip.2cm \noindent [6] R. Gorenflo and F. Mainardi, The Mittag-Leffler function in the Riemann-Liouville fractional calculus, In: A.A. Kilbas (ed.) Boundary Value Problems, Special Functions and Fractional Calculus, Minsk, 1996, pp. 215-225.

\vskip.2cm \noindent [7] R. Gorenflo, A.A. Kilbas, and S.V. Rogosin, On the generalized Mittag-Leffler type function, Integral Transforms and Special Functions, 7(1998), 215-224.

\vskip.2cm \noindent [8] R. Gorenflo, J. Loutchko, and Yuri Luchko, Computation of the Mittag-Leffler function and its derivatives, Fractional Calculus and Applied Analysis, 5(2002), 491-518.

\vskip.2cm \noindent [9] H.J. Haubold and  A.M. Mathai, The fractional kinetic equation and thermonuclear functions, Astrophysics and Space Science, 273(2000), 53-63;

\vskip.2cm \noindent [10] H.J. Haubold and D. Kumar, Fusion yield: Guderley model and Tsallis statistics, Journal of Plasma Physics, doi: 10.1017/S0022377810000590.

\vskip.2cm \noindent [11] H.J. Haubold, A.M. Mathai, and R.K. Saxena, Further solutions of fractional reaction-diffusion equations in terms of the H-function, Journal of Computational and Applied Mathematics, 235(2011), 1311-1316.

\vskip.2cm \noindent [12] K. Jayakumar, On Mittag-Leffler process, Mathematical and Computer Modelling, 37(2003), 1427-1434.

\vskip.2cm \noindent [13] K. Jayakumar and R.P. Suresh, Mittag-Leffler distribution, Journal of the Indian Society of Probability and Statistics, 7(2003), 51-71.

\vskip.2cm \noindent [14] A.A. Kilbas, Fractional calculus of the generalized Wright function, Fractional Calculus and Applied Analysis, 8(2005), 113-126.

\vskip.2cm \noindent [15] A.A. Kilbas and M. Saigo, On Mittag-Leffler type functions, fractional calculus operators and solution of integral equations, Integral Transforms and Special Functions, 4(1996), 335-370.

\vskip.2cm \noindent [16] V. Kiryakova, Multiple (multiindex) Mittag-Leffler functions and relations to generalized fractional calculus, Journal of Computational and Applied Mathematics, 118(2000), 214-259;

\vskip.2cm \noindent [17] S.C. Lim and L.P. Teo, Analytic and asymptotic properties of multivariate generalized Linnik's probability densities, arXiv.org/abs/0903.5344 [math.PR] 30 March 2009.

\vskip.2cm \noindent [18] Gwo Dong Lin, On the Mittag-Leffler distribution, Journal of Statistical Planning and Inference, 74(1998), 1-9.

\vskip.2cm \noindent [19] F. Mainardi and G. Pagnini, Mellin-Barnes integrals for stable distributions and their convolutions, Fractional Calculus and Applied Analysis, 11(2008), 443-456;

\vskip.2cm \noindent [20] F. Mainardi, Fractional Calculus and Waves in Linear Viscoelasticity: An Introduction to Mathematical Models, Imperial College Press, London, 2010.

\vskip.2cm \noindent [21] A.M. Mathai, A Handbook of Generalized Special Functions for Statistical and Physical Sciences, Clarendon Press, Oxford, 1993.

\vskip.2cm \noindent [22] A.M. Mathai, Jacobians of Matrix Transformations and Functions of Matrix Argument, World Scientific Publishers, New York, 1997.

\vskip.2cm \noindent [23] A.M. Mathai, An Introduction to Geometrical Probabilities: Distributional Aspects with Applications, Gordon and Breach, New York, 1999.

\vskip.2cm \noindent [24] A.M. Mathai, Pathways to matrix-variate gamma and normal densities, Linear Algebra and Its Applications, 396(2005), 317-328.

\vskip.2cm \noindent [25] A.M. Mathai, Random volumes under a general matrix-variate model, Linear Algebra and Its Applications, 425(2007), 162-170.

\vskip.2cm \noindent [26] A.M. Mathai, Some properties of Mittag-Leffler functions and matrix-variate analogues: A statistical perspective, Fractional Calculus and Applied Analysis, 13(2010), 113-132.

\vskip.2cm \noindent [27] A.M. Mathai and H.J. Haubold, On generalized entropy measures and pathways, Physica A, 385(2007), 493-500;

\vskip.2cm \noindent [28] A.M. Mathai and H.J. Haubold, Mittag-Leffler functions to pathway model to Tsallis statistics, Integral Transforms and Special Functions, 21(2010), 867-875;

\vskip.2cm \noindent [29] A.M. Mathai, H.J. Haubold, and C. Tsallis, Pathway model and nonextensive statistical mechanics, arXiv.org/abs/1010.4597 [cond-mat.stat-mech] 21 October 2010.

\vskip.2cm \noindent [30] A.M. Mathai and H.J. Haubold, Special Functions for Applied Scientists, Springer, New York, 2008.

\vskip.2cm \noindent [31] A.M. Mathai and S. Provost, Quadratic Forms in Random Variables: Theory and Applications, Marcel Dekker, New York, 1992.

\vskip.2cm \noindent [32] A.M. Mathai, S. Provost, and T. Hayakawa, Bilinear Forms and Zonal Polynomials, Springer, New York, 1995.

\vskip.2cm \noindent [33] A.M. Mathai and P.N. Rathie, Basic Concepts in Information Theory and Statistics: Axiomatic Foundations and Applications, Wiley Halsted, New York and Wiley Eastern, New Delhi, 1975.

\vskip.2cm \noindent [34] A.M. Mathai, R.K. Saxena, and H.J. Haubold, A certain class of Laplace transforms with application in reaction and reaction-diffusion equations, Astrophysics and Space Science, 305(2006), 283-288;

\vskip.2cm \noindent [35] A.M. Mathai, R.K. Saxena, and H.J. Haubold, The H-function: Theory and Applications, Springer, New York, 2010.

\vskip.2cm \noindent [36] A.G. Pakes, Mixture representation for symmetric generalized Linnik laws, Statistics and Probability Letters, 37(1998), 213-221.

\vskip.2cm \noindent [37] R.N. Pillai, On Mittag-Leffler functions and related distributions, Annals of the Institute of Statistical Mathematics, 42(1990), 157-161.

\vskip.2cm \noindent [38] R.N. Pillai and K. Jayakumar, Discrete Mittag-Leffler distributions, Statistics and Probability Letters, 23(1995), 271-274.

\vskip.2cm \noindent [39] R.K. Saxena, A.M. Mathai, and H.J. Haubold, Solutions of certain fractional kinetic equations and a fractional diffusion equation, Journal of Mathematical Physics, 51(2010), 103506;

\vskip.2cm \noindent [40] J. Tenreiro Machado, V. Kiryakova, and F. Mainardi, Recent history of fractional calculus, Communications in Nonlinear Science and Numerical Simulation, 16(2011), 1140-1153.

\vskip.2cm \noindent [41] C. Tsallis, Possible generalization of Boltzmann-Gibbs statistics, Journal of Statistical Physics, 52(1988), 479-487;

\vskip.2cm \noindent [42] C. Tsallis, What should statistical mechanics satisfy to reflect nature?, Physica D, 193(2004), 3-34;

\vskip.2cm \noindent [43] C. Tsallis, Nonadditive entropy and nonextensive statistical mechanics - An overview after 20 years, Brazilian Journal of Physics, 39(2009), 337-356;

\vskip.2cm \noindent [44] C. Tsallis, Introduction to Nonextensive Statistical Mechanics: Approaching a Complex World, Springer, New York, 2009.

\vskip.2cm \noindent [45] E.M. Wright, The asymptotic expansion of the generalized hypergeometric function, Proceedings of the London Mathematical Society, 46(1940), 389-408.

\vskip.2cm \noindent [46] G. Cottone, M. Di Paola, and R. Metzler, Fractional calculus approach to the statistical characterization of random variables and vectors, Physica A, 389(2010), 909-920.

\vskip.3cm

[1] Centre for Mathematical Sciences Pala Campus\\
Arunapuram P.O., Pala, Kerala-686574, India and \\
Department of Mathematics and Statistics, McGill University, Canada, H3A2K6\\
Email: mathai@math.mcgill.ca\\

\vskip.3cm and\\

[2] Office for Outer Space Affairs, United Nations\\
P.O. Box 500, Vienna International Centre, A-1400 Vienna, Austria and\\
Centre for Mathematical Sciences Pala Campus, Arunapuram P.O., Pala, Kerala-686574, India\\
Email: hans.haubold@unvienna.org\\

\end{document}